\documentclass[12pt,a4paper]{article}
\usepackage{epsfig}
\begin{document}
\textwidth=135mm
 \textheight=200mm
\begin{center}
{\bfseries  Hadron calorimeter module prototype for baryonic matter studies at Nuclotron}
\vskip 5mm
O.P.~Gavrishchuk$^{\dag,\ddag}$,  V.P.~Ladygin$^{\dag,}$\footnote{Corresponding author: vladygin@jinr.ru}, 
Yu.P.~Petukhov$^\dag$ and  S.Ya.~Sychkov$^\dag$
\vskip 5mm
{\small {\it $^\dag$ Joint Institute for
Nuclear Research, 141980 Dubna, Russia}} \\
{\small {\it $^\ddag$E-mail: Oleg.gavrishchuk@cern.ch }}\\
\end{center}
\vskip 5mm
\centerline{\bf Abstract}
The prototype of the hadron calorimeter module consisting of 66 scintillator/lead layers 
with the 15$\times$15~cm$^2$ cross section and $\sim$5 nuclear interaction lengths has been designed and produced for the zero degree calorimeter of the BM@N experiment.
The prototype has been tested  with high energy muon beam of the  U-70 accelerator at IHEP.
The results of the beam test for different types of photo multipliers and light guides are presented. 
The results of the Monte-Carlo simulation of the calorimeter response and energy resolution 
are presented for the 2-16 GeV protons.
\vskip 10mm
\section{\label{sec:intro}Introduction}
The study of the dense baryonic matter at Nuclotron (BM@N project) \cite{bmn_CDR} 
is proposed as a first stage in the heavy-ion  program at NICA \cite{nica}. 
The major direction in the research program of BM@N project is the production of 
strange matter in heavy-ion collisions at beam energies between 2 and 6 A$\cdot$GeV \cite{bmn_PoS}. 
The other topics of the experimental program can be related with the study of in-medium effects for 
strange particles decaying in hadronic modes \cite{brat1}, hard probes and correlations
\cite{vasiliev_npps2011}, soft photons and neutral mesons \cite{svd2}, polarization effects
\cite{bmn_dspin2013} etc. For these purposes an experimental setup will be installed 
at the 6V beamline in the fixed-target hall of the Nuclotron.   
The first results  with the relativistic deuteron \cite{terekhin_PoS} and carbon 
\cite{piyadin_C12} beams  demonstrated the feasibility of these studies with light nuclei.

An important direction of the relativistic heavy-ion collisions studies is the measurements of different
observables as a function of the centrality parameter. The concept of centrality is based on the use of the impact parameter value which possible to define as a distance between the centres of the colliding nuclei in the plane, perpendicular to the beam direction. Since the impact parameter is a non-measurable value, it is possible to use such characteristic of the collision, as the part of a total energy which is carried away by the spectators: protons, neutrons and nuclear fragments. Most of spectators has a small transverse momentum and is emitted in a narrow cone near the primary beam direction. To measure the  spectators  energy the calorimeter at zero angle, so-called Zero Degree Calorimeter (ZDC), is usually used.
Another purpose of ZDC is the central events selection at the trigger level during data taking. 

\begin{figure}[hbtp]
 \centering
  \resizebox{12cm}{!}{\includegraphics{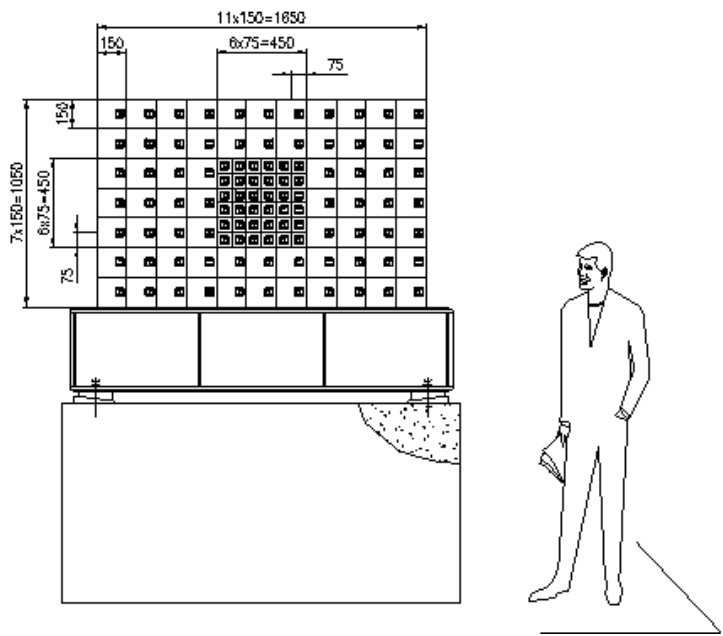}}
\caption{Transversal view of ZDC consisting of 104 HCAL modules. Central and peripheral  parts and  consist of  6$\times$6 modules and 68 modules with cross section of 7.5$\times$7.5 cm$^2$ and  
15$\times$15 cm$^2$, respectively.}
\label{fig:fig1}
\end{figure}
 
Main ZDC option for the first stage of BM@N experiment \cite{bmn_CDR} 
will be based on the scintillator/lead sandwich type of the hadron calorimeter (HCAL) 
used in WA98  \cite{WA98_cal} and COMPASS \cite{Compass_cal} experiments at CERN.
Central part of ZDC  consists of an array of 6$\times$6 modules with the cross section
of 7.5$\times$7.5 cm$^2$ each,
while peripheral part contains 68 modules of 15$\times$15 cm$^2$ 
transversal size.  Such granularity allows one to operate with the
nuclear beam intensities up to 1 MHz, which is sufficient for the first stage of the experiment.
For the beam intensity of 10 MHz it will be necessary to remove several central modules.
The schematic view of the transverse cross section of ZDC for BM@N experiment is presented in 
Fig.\ref{fig:fig1}.
The HCAL module \cite{Compass_cal} consists  of 80 scintillator/lead layers with the sampling structure
of 1/4. This sampling is optimal to obtain the $e/\pi$ and $e/$p ratios close to 1 and the
good energy resolution for all particles in the energy range of 2$\div$100 GeV with $\sigma_E$/$E$ 
$\sim$ 50\%/$\sqrt{E(GeV)}$. The results of the experiment at Nuclotron with the use of $^6Li$ beam
shown the similar  energy resolution for nucleons \cite{balandin}. 
However, optimization of the module design for Nuclotron energies is required.

In this paper the results of the Monte-Carlo simulation on the optimization  of the 
scintillator/lead hadronic calorimeter for Nuclotron energies as well as the systematic 
studies of the HCAL module for the peripheral part of ZDC with muon beam are presented. 

\section{\label{sec:MC}Monte-Carlo simulation for BM@N ZDC}

\begin{figure}[hbtp]
\begin{minipage}[t]{0.47\textwidth}
 \centering
  \resizebox{6.5cm}{!}{\includegraphics{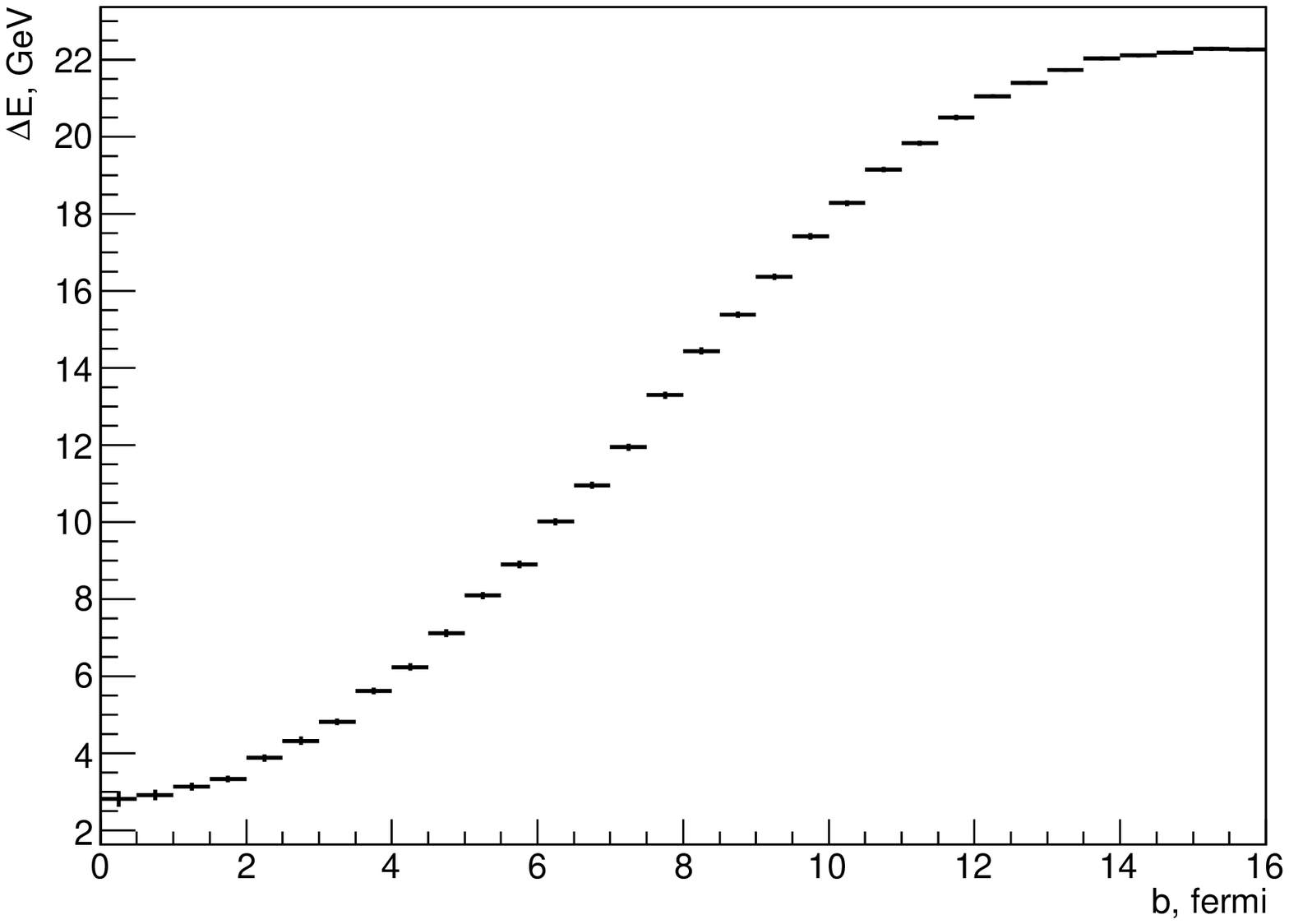}}
\end{minipage}\hfill
\begin{minipage}[t]{0.47\textwidth}
 \centering
  \resizebox{6.5cm}{!}{\includegraphics{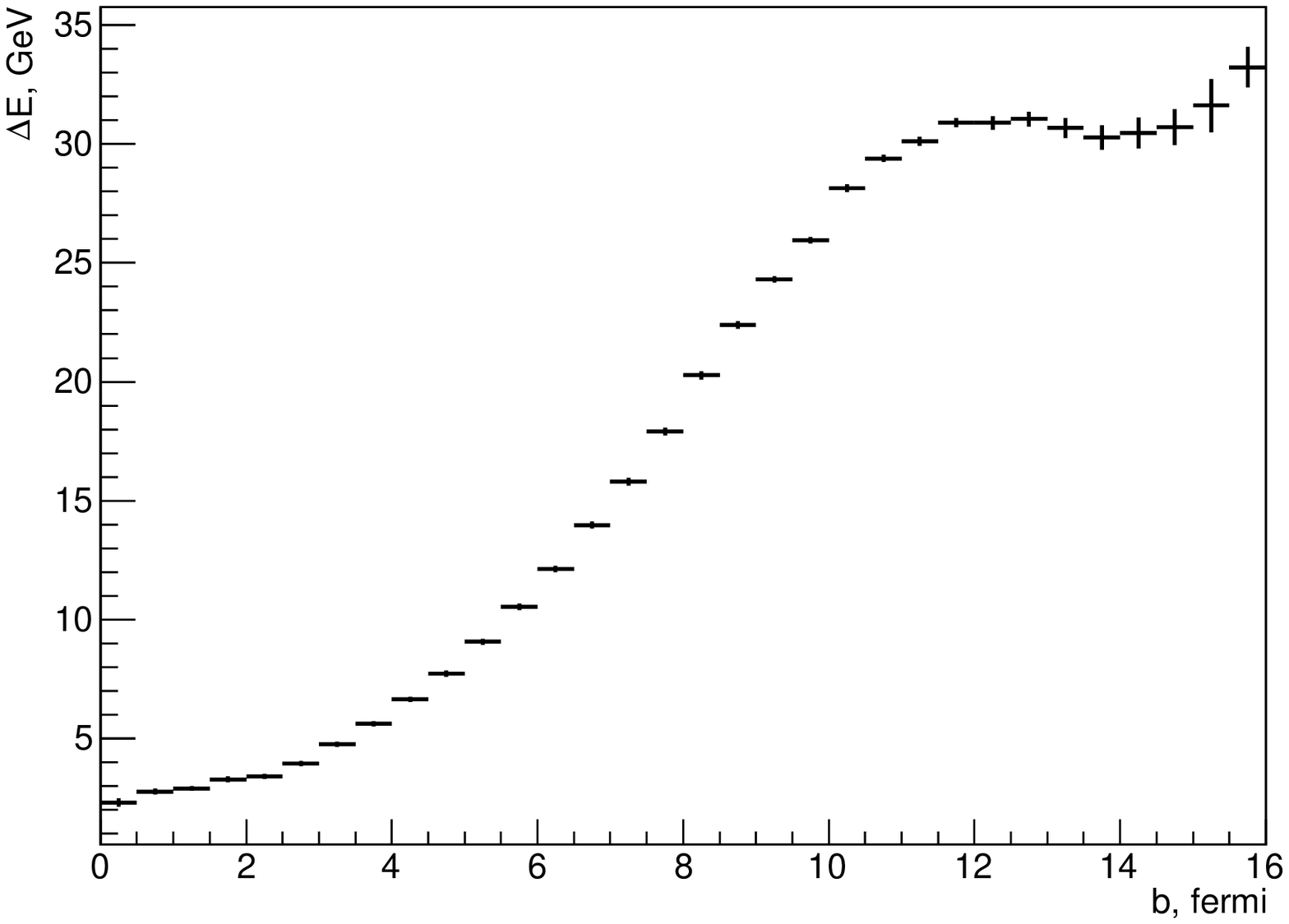}}
 \end{minipage}
\caption{The total energy $\Delta E$  deposited in ZDC as a function of impact parameter  $b$ for UrQMD \cite{urqmd} (left) and LAQGSM \cite{laqgsm} minimal bias events at 4 GeV/nucleon. }
\label{fig:fig2}
\end{figure}

\begin{figure}[hbtp]
 \centering
  \resizebox{10cm}{!}{\includegraphics{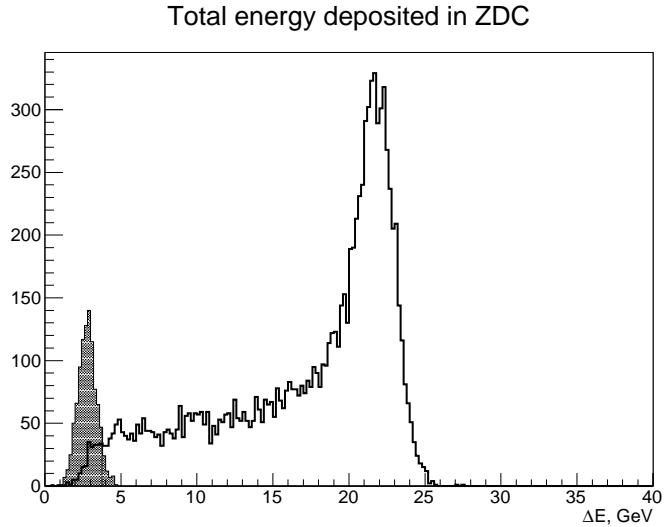}}
\caption{The energy deposition in ZDC for the central and minimum bias UrQMD \cite{urqmd} events shown by hatched and open histograms, respectively.}
\label{fig:fig3}
\end{figure}

One of the main parameters of ZDC is the average part of the total spectators energy
registered in the calorimeter related to the total spectators energy. With this part increasing
the events classification by centrality appear more reliable. Other important ZDC parameter 
is the energy resolution. The provided error in the measurement of the
total spectators energy should be less, than average fluctuations of the spectators energy
for the central collisions. Thus, angular acceptance (actually cross-section size) and energy
resolution should provide high efficiency of events selection on the basis of the centrality.

The simulation of the ZDC response  has been performed using Au+Au central and minimum bias
UrQMD \cite{urqmd} and LAQGSM \cite{laqgsm} events at 4.0 GeV/nucleon and BM@N software framework described in 
ref.\cite{bmn_dspin2013}. The transverse size of ZDC was taken as shown in Fig.\ref{fig:fig1}.
The position of ZDC was taken as 10 m from the target point.  
The fraction of the total energy deposited in ZDC is about 0.6 for the maximal magnetic field 
in the BM@N magnet. This fraction will increase with the magnetic field decreasing.  
The dependences of the deposited in ZDC total energy $\Delta E$ on the impact parameter $b$  for the minimum bias UrQMD \cite{urqmd} and LAQGSM \cite{laqgsm} events are shown in the left and right panels of Fig. \ref{fig:fig2}, respectively. The correlation of the  total energy $\Delta E$  and impact parameter $b$ clearly seen in Fig. \ref{fig:fig2} allows to use ZDC information to estimate the centrality of the collisions. One can see, that there is no difference for the
central and semi-central events  for  UrQMD \cite{urqmd} and LAQGSM \cite{laqgsm} models.
However,  LAQGSM \cite{laqgsm} gives larger energy deposition for the minimal bias events due to
taking into account nuclear fragments. 

The energy deposition in ZDC for the 1000 central and 10000 minimum bias UrQMD \cite{urqmd} events is shown in Fig.\ref{fig:fig3} by the hatched and open histograms, respectively. The average value of the deposited energy in ZDC is $\sim$3 GeV and $\sim$22 GeV for the central and 
minimum bias events, respectively.  Therefore, the signal from ZDC can be used  also for the trigger
purposes to select the events within fixed range of the centrality.

\begin{figure}[hbt!]
 \centering
  \resizebox{10cm}{!}{\includegraphics{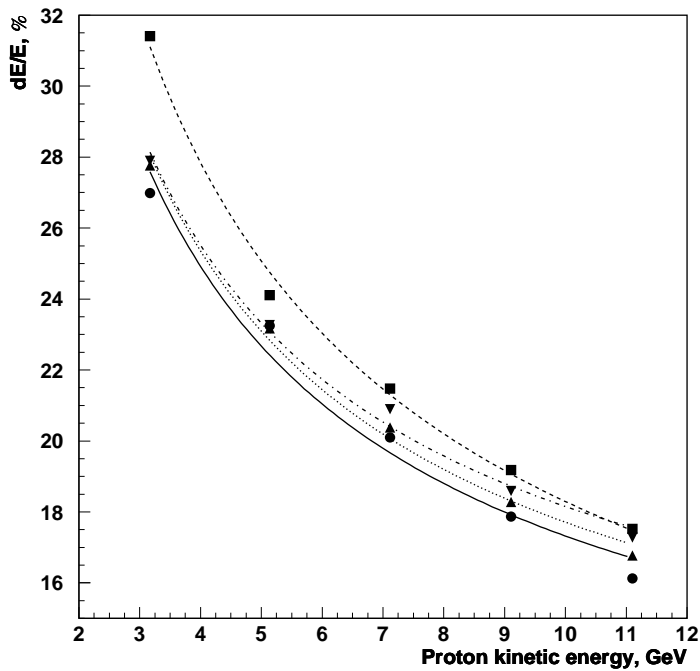}}
\caption{ZDC energy resolution $\sigma_E$/$E$ as a function of the  proton kinetic energy  $T$   for different   scintillator/lead   sampling structures.   
The symbols and lines corresponds to the following number of scintillator/lead  layers $n$ and scintillator thickness $l_s$: 
circles and solid line - $n=65$ and $l_s=4$mm, triangles and dotted line -
$n=65$ and $l_s=5$mm, rotated triangles and dash-dotted line - $n=49$ and $l_s=10$mm, 
squares and dashed line - $n=78$ and $l_s=2.5$mm, respectively.
The lines are the results of the interpolation by the function   $a$/$\sqrt{E(GeV)}$ 
$\oplus b$.}
\label{fig:fig4}
\end{figure}

ZDC energy resolution plays 
ZDC energy resolution $\sigma_E$/$E$ as a function of the  proton kinetic energy  $T$ 
has been studied for different calorimeter scintillator/lead  sampling structures. The lead absorber plate thickness was  fixed to 10 mm, while the number of scintillator/lead  layers $n$ and scintillator
thickness $l_s$ were changed. The results for the $n=78$ and $l_s=2.5$mm, 
$n=65$ and $l_s=4$mm, $n=65$ and $l_s=5$mm, $n=49$ and $l_s=10$mm are shown 
in Fig.\ref{fig:fig4} by the solid squares, 
circles, triangles  and rotated triangles, respectively. The lines are the results of the interpolation by the function   $a$/$\sqrt{E(GeV)}$ $\oplus b$. 
The results of the interpolation for different  calorimeter scintillator/lead  sampling structures
are summarized in Table 1.

\begin{table}[hbt!]
\label{table:zdc-res}
\caption{The simulation results for ZDC energy resolution for different
calorimeter scintillator/lead  sampling structures. Here $n$ and $l_s$ are the number
of scintillator/lead  layers and scintillator thickness, respectively. The lead absorber plate thickness
is fixed to 10 mm. The parameters $a$ and $b$
are obtained from the interpolation of the $\sigma_E$/$E$ energy dependence presented in Fig.\ref{fig:fig4} by the function
$a$/$\sqrt{E(GeV)}$ $\oplus b$.} 
\vspace{5mm}
\centering
\begin{tabular}{|c|c|c|}
\hline
Scintillator/ &    $a$, &  $b$,   \\
lead sampling &    [\%] &  [\%]   \\
\hline
$n=78$, $l_s=2.5$mm &  54.2 & 6.4  \\
$n=65$, $l_s=4$mm   &  46.3 & 9.3  \\
$n=65$, $l_s=5$mm   &  46.9 & 9.7  \\
$n=49$, $l_s=10$mm  &  46.3 & 10.7  \\
\hline
\end{tabular}
\end{table}

The results indicate that the achieved energy resolution is better than 50\%/$\sqrt{E(GeV)}$
for the scintillator thickness $l_s\ge$4 mm. However, the homogeneity (parameter $b$) 
decreases with the increasing of the scintillator plates thickness.  
One can see, that the sampling structures  $n=65$, $l_s=4$mm or $n=65$, $l_s=5$mm provide 
optimal values of the $a$ and $b$ parameters in the considered energy range.   
These results were taken into account to design of the HCAL module prototype.

\section{\label{sec:module} ZDC module prototype}

\begin{figure}[hbtp]
 \centering
  \resizebox{12cm}{!}{\includegraphics{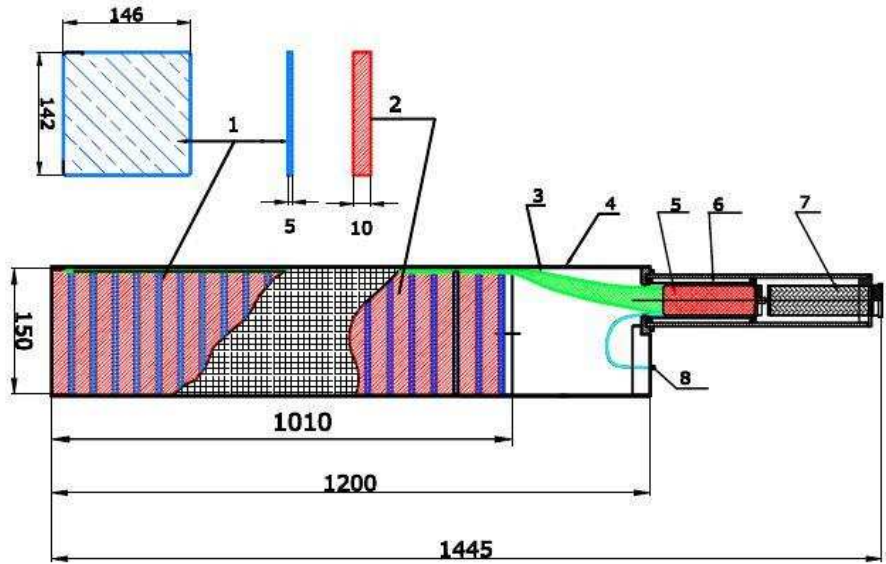}}
\caption{ZDC module prototype design. 1,2 - 66 scintillator/lead layers, 3 - WLS,  4 - iron housing, 5 - PMT, 6 - $\mu$-metal screen,  7 - PM high voltage base,  8 - fiber with an optical connector.}
\label{fig:fig5}
\end{figure}

The HCAL module  design is shown in Fig.\ref{fig:fig5}.   
The module transverse  size is  equal to 15$\times$ 15 cm$^2$, 
iron housing  length is equal 120 cm and total length with photo-multiplier tube  (PMT)
is equal 145 cm.The  module components are  placed in  the light protection box which is as well module housing. 
The HCAL module prototype sampling structure consists of 66 scintillator/lead layers with thickness 4 mm and 10 mm, respectively. 
Module prototype  absorption  length is about 5 interaction nuclear lengths ($\lambda_{int}$). 
In this case ZDC energy resolution for hadrons 2-30 GeV  is
$\sigma_E$/$E$=46.3\%/$\sqrt{E(GeV)}$ 
$\oplus$9.3\%  according to the Monte-Carlo simulation results shown in Fig.\ref{fig:fig4}.   

The light from scintillator plates  is transmitted from one side with a fast wavelength shifter (WLS)  to a PMT placed inside the $\mu$-metal screen and coupled with an high voltage base. 
The light emitted diode (LED) radiating a light with the wave length of  460 nm is used for 
the tests.
The light from LED  is transmitted to the photo-cathode of PMT via 1 mm diameter optical fiber and the optical connectors located at the rear side of the HCAL module prototype.
The LED signal can be  distributed among all calorimeter modules.

\begin{figure}[hbtp]
\begin{minipage}[t]{0.47\textwidth}
 \centering
  \resizebox{6.0cm}{!}{\includegraphics{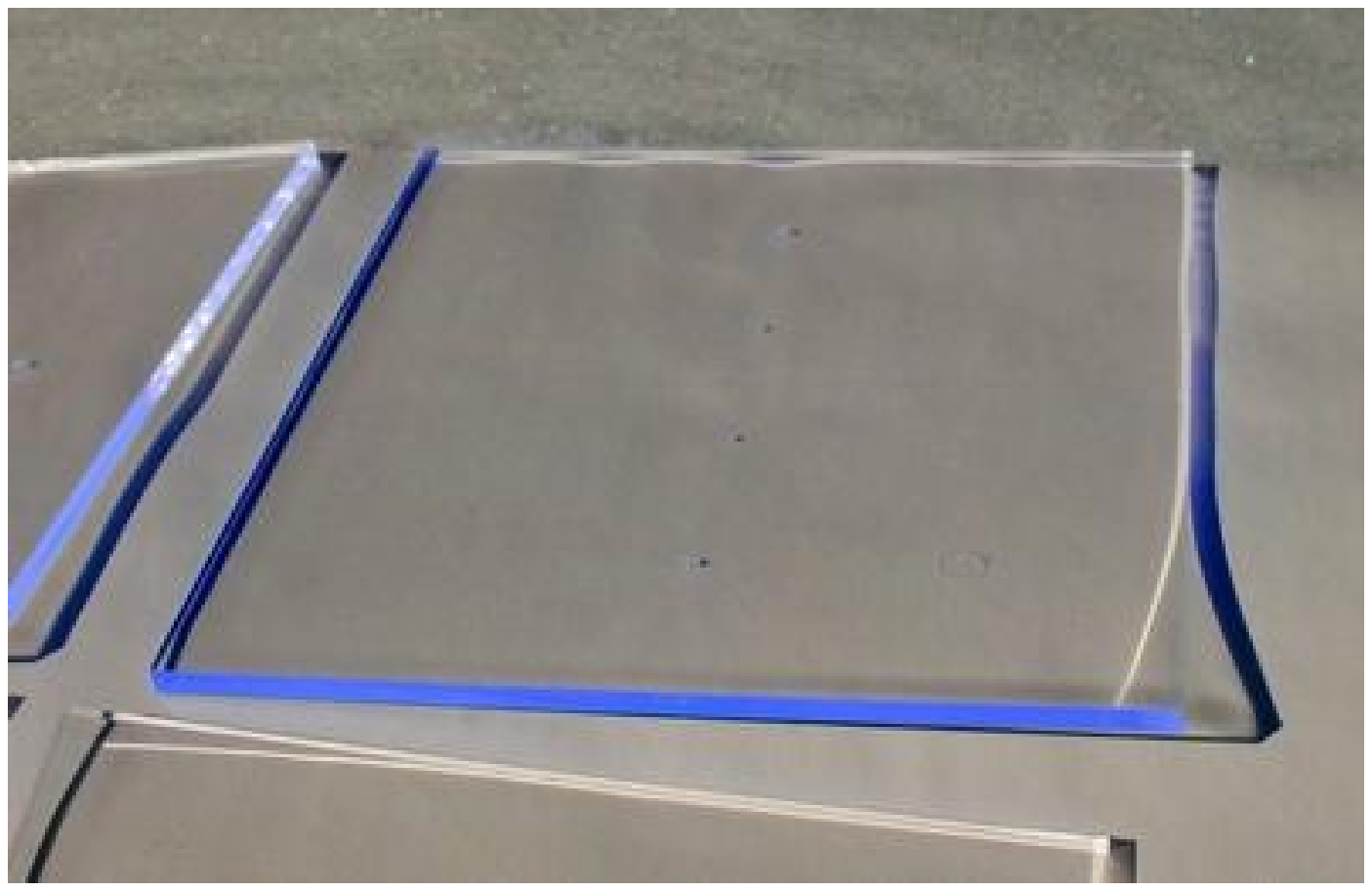}}
\end{minipage}\hfill
\begin{minipage}[t]{0.47\textwidth}
 \centering
  \resizebox{6.0cm}{!}{\includegraphics{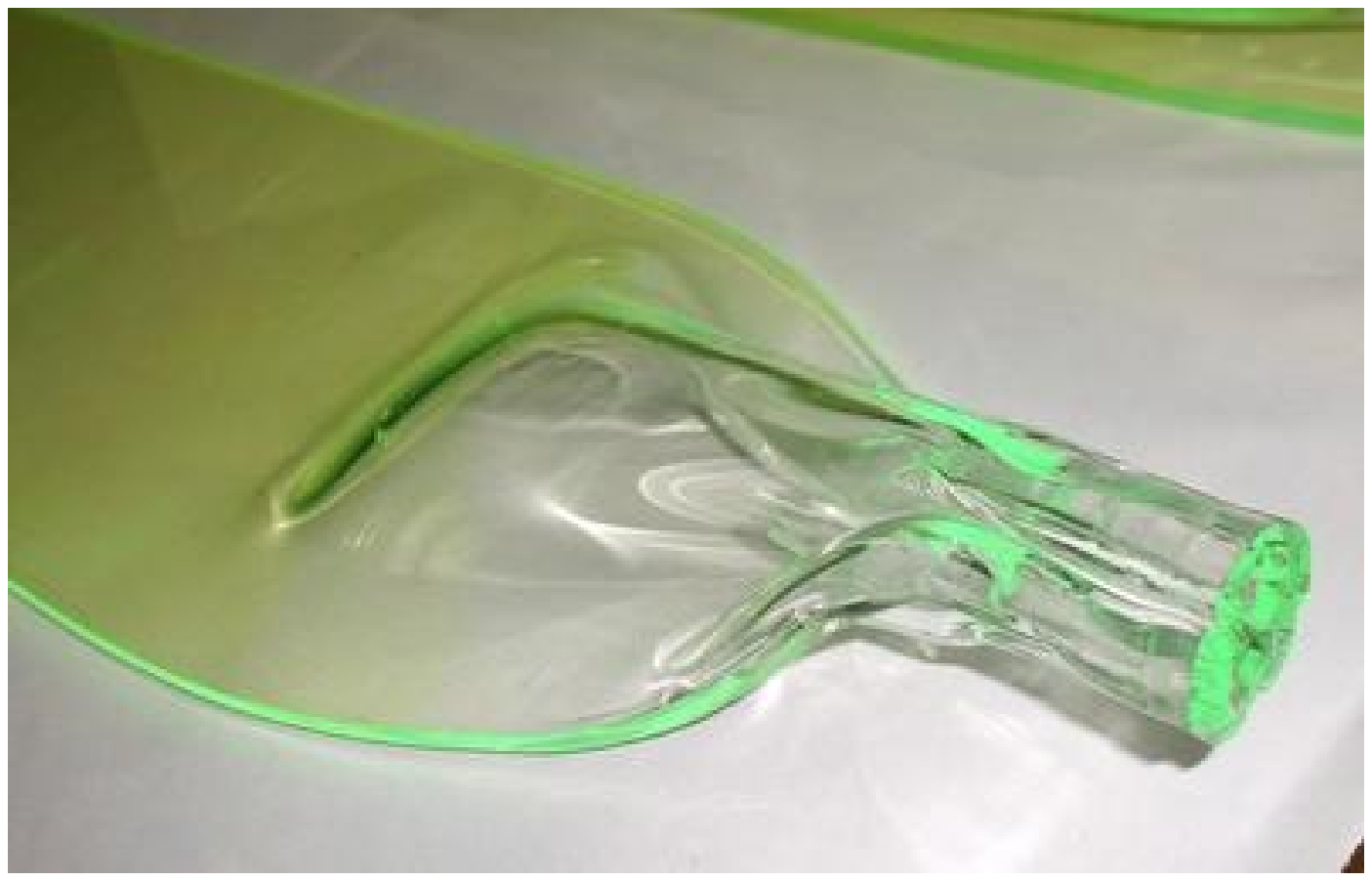}}
 \end{minipage}
\caption{The  components of the HCAL module prototype: scintillation plates (left) and light guide with  wave length
shifter (right).}
\label{fig:fig6}
\end{figure}

\begin{figure}[hbtp]
 \centering
  \resizebox{12cm}{!}{\includegraphics{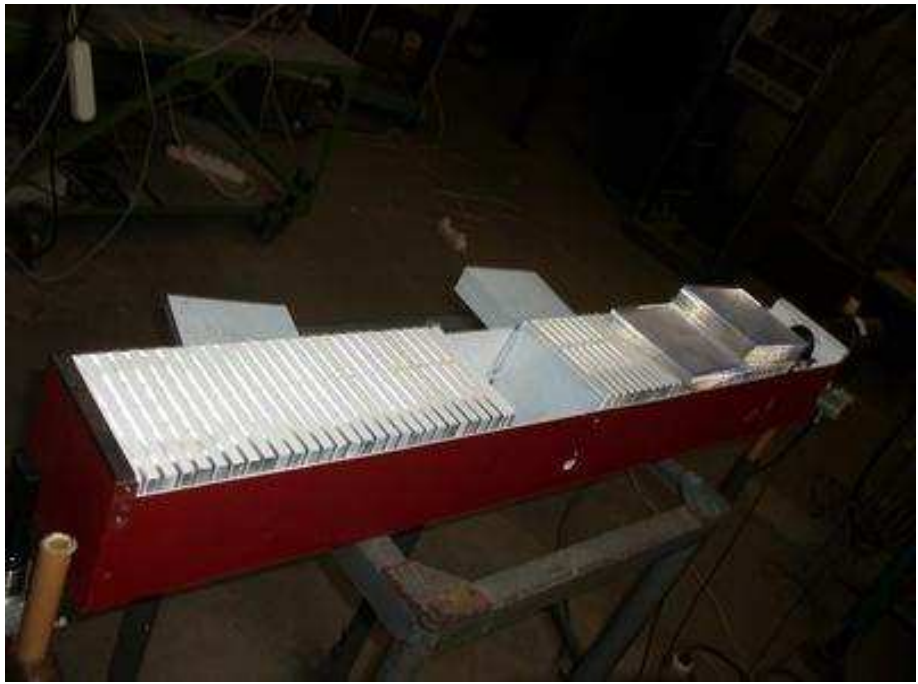}}
\caption{Inside view of the calorimeter module prototype.}
\label{fig:fig7}
\end{figure} 

The parts of WA98 hadron calorimeter \cite{WA98_cal} will be used for assembling of BM@N ZDC \cite{bmn_CDR}. 
Unfortunately, the light output from scintillators used in the WA98 calorimeter is decreased due to radiation damage. Also the light outputs for the scintillators taken from
the centre and periphery of the WA98 calorimeter are different. The measurements were
performed using light guide 20$\times$150 mm$^2$ and FEU-85 PMT. The LeCroy-2249A ADC has
been previously calibrated using photo-diode, the bin width is equals to 0.46 p.e.  
The light output is decreased for $\sim$20\% and   $\sim$40\% for the periphery and 
centre of the WA98 ZDC \cite{bmn_CDR}. 

Therefore, for the HCAL module prototype new scintillator
plates were  manufactured at IHEP using the method of the injection
molding under pressure from the granulated polystyrene PSM-115.
The additives composition is P-terphenyl of 1.5 \% and POPOP of 0.05\%.
Scintillator of this type have been successfully used in calorimeters for the last 20 years. 
It has a high radiation tolerance - about 10 MRad, a good light output  
(of $\sim$60\% from  the antracene light output), a fast decay time $\sim$1.5 ns, a good  transparency for blue light (wave length $\sim$420 nm)  with attenuation  length of about 30-60 cm.

\begin{figure}[hbtp]
\begin{minipage}[t]{0.47\textwidth}
 \centering
  \resizebox{6.0cm}{!}{\includegraphics{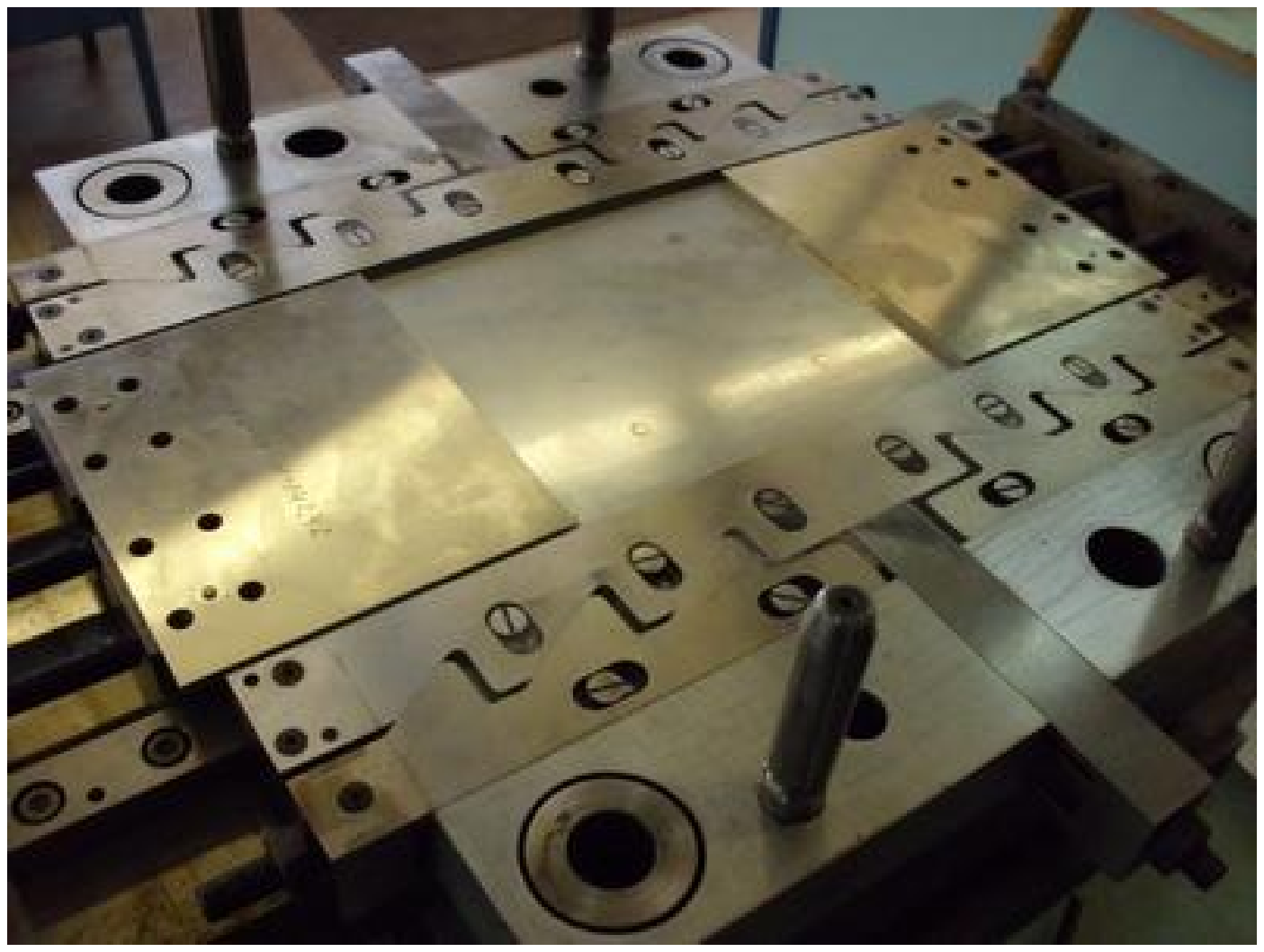}}
\end{minipage}\hfill
\begin{minipage}[t]{0.47\textwidth}
 \centering
  \resizebox{6.0cm}{!}{\includegraphics{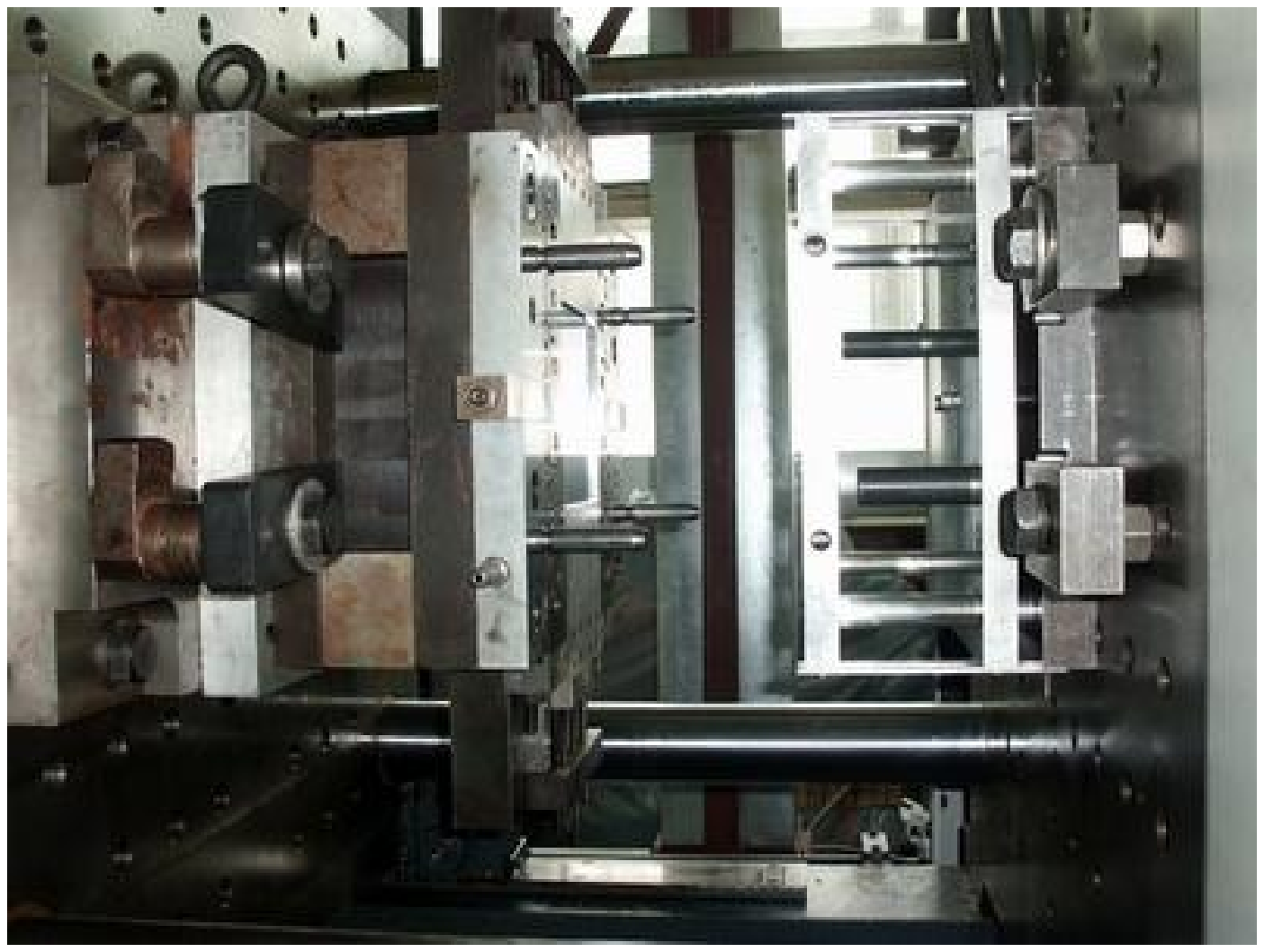}}
 \end{minipage}
\caption{The matrix for the scintillator production (left) and 
     and  matrix installed on the injection machine (right).}
\label{fig:fig12}
\end{figure}

The injection molding under pressure technology requires the use of the special matrix  for scintillation plates production with given sizes. This matrix should be installed on a special injection machine with heating, extrusion and pressure systems. 
There is an cavity for the scintillator plates of given size at the centre of the matrix. The cavity  shape have variable transverse  size with the fixed thickness of 4 mm. The use of this 
matrix  allows to  change the scintillation plates dimension up to 160$\times$300 mm$^2$.
The produced scintillator  plates transfer size and thickness are equal to 142$\times$146 mm$^2$  and 4 mm, respectively.  The matrix for the scintillator production as well as 
matrix with the injection machine are shown in Fig.\ref{fig:fig12}.

WLS task is to absorb the   blue light with a wave length $\sim$420 nm  
produced in the scintillator and radiate green light spectra with a wave length 
$\sim$520 nm, which is optimal  for the PMTs with bi-alkaline  photocathodes.  
The WLS is manufactured from the optical organic glass (PMMA — polymethylmethacrylate) 
with the surface painted by coumarin K-7 dissolved in the ethanol at 60$^\circ$ C. The
deepness of K-7 diffusion has varied in range of 5-15 $\mu$m. It allows one to obtain the good
light collection uniformity along the WLS length (1 m) with the attenuation length larger
than 3 m. The WLS has a length of 120 cm, width  and thickness are equal to 12 cm and  0.4 cm, respectively.   The scintillation plates and light guide with  wave length shifter used
for the HCAL module prototype are shown in Fig.\ref{fig:fig6}.

The absorber was made of lead plates 10 mm thick and 142$\times$146 mm$^2$ transverse sizes.  These lead plates were reused from  WA98  hadron calorimeter  \cite{WA98_cal}.  The absorber was made from lead with 2\% antimony to increase the hardness of the material. 
%Lead plates were manufactured in JINR (1994) by  the molding in the form of given size.  
The light shielding box for module was made from steel sheets with a thickness of 1.5 mm.  
Scintillation plates   were not wrapped in reflective coating,  while the lead plates and module housing inside  were painted by white color to obtain diffusion light reflection. 
The mirror from Al-mylar was embedded at the bottom of the housing to reflect the light from the bottom  edge. The inside view of the HCAL module prototype is shown in Fig.\ref{fig:fig7}.

\section{\label{sec:btc} Beam test  conditions.}

HCAL module prototype test was performed using muon beam with an energy of $\sim$10 GeV.
The goal of the beam test was to measure the  light yield  and timing resolution  for
the minimum ionizing particles (MIPs) for the prototype for different types of PMT and WLS,
as well as  its transverse uniformity. 

\begin{figure}[hbtp]
 \centering
  \resizebox{12cm}{!}{\includegraphics{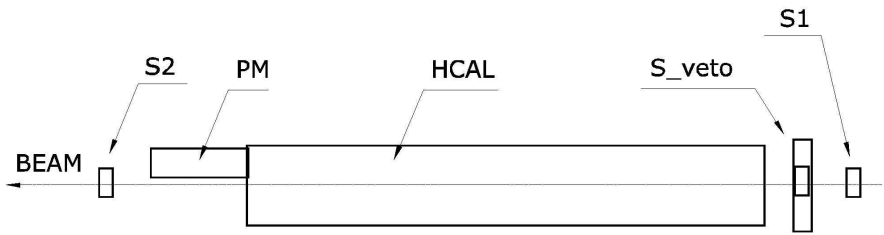}}
\caption{The layout of the  beam test area. HCAL is the hadron calorimeter module prototype; 
$S_1$, $S_2$, $S_{veto}$  is the scintillation counters for trigger logic and definition the beam position, PM is the photomultiplier.}
\label{fig:fig8}
\end{figure}

HCAL module prototype was installed  in the beam test area as shown in Fig.\ref{fig:fig8}.
Muons were generated by the secondary hadrons produced from the interaction of the proton beam with the internal nuclear target at accelerator U-70 and passed throughout the  reinforced concrete protection. The muon beam has an energy $\sim$10 GeV with intensity up to 10$^5$/cm$^2$ per spill.  
High energy muon beam passes through the HCAL module prototype parallel to its axis. 

The signals from three scintillation counters $S_1$, S$_2$ and $S_{veto}$  
were used for the trigger purposes. The scintillators were coupled with the fast R3478 PMTs to obtain a good timing resolution of the trigger logic. 
Two of them, $S_1$ and S$_2$, were installed in front and back of the prototype, respectively.  
Transverse dimension and thickness of their scintillators were 3$\times$3 cm$^2$  and 2 cm, respectively. The  third counter $S_{veto}$  with the 3$\times$3 cm$^2$ hole 
in the centre of the  scintillator was also placed in front of the HCAL module prototype. 
The signals from PMT anodes were fed to a pulse shaper with constant threshold of $\sim$50 mV through 50 ohms splitter. 
The signals coincidence from $S_1$ and S$_2$ counters  defined the impact position and 
transverse size of the muon beam. The use 
of the  $S_{veto}$ counter signal in the trigger logic allowed to avoid the  multiple particles  hitting in the module prototype.
The LeCroy2249A and LeCroy2228A modules were used to store the charge and timing information
from the scintillation counters and PMT of the HCAL module, respectively.

\begin{figure}[hbtp]
\begin{minipage}[t]{0.47\textwidth}
 \centering
  \resizebox{6.5cm}{!}{\includegraphics{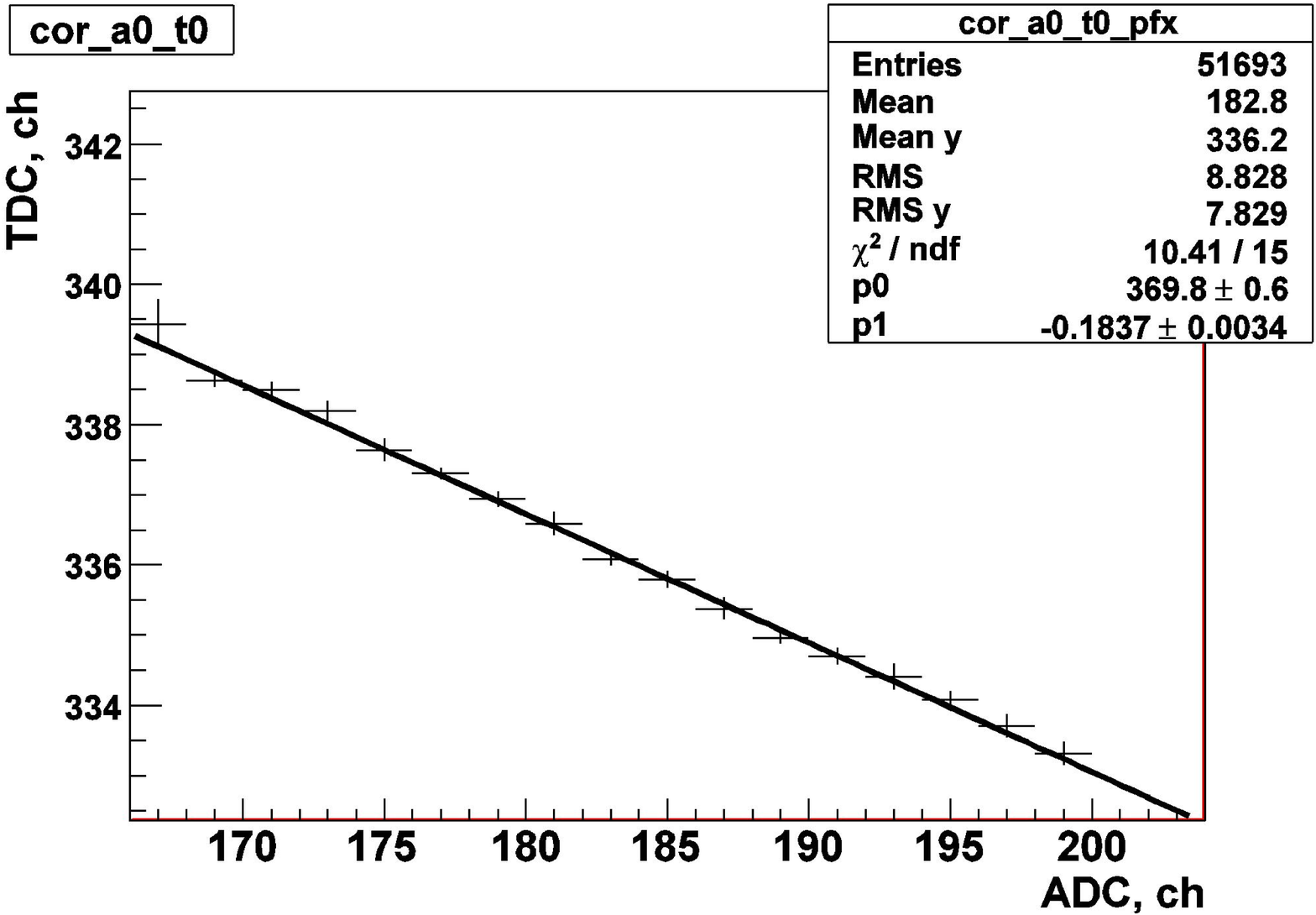}}
\end{minipage}\hfill
\begin{minipage}[t]{0.47\textwidth}
 \centering
  \resizebox{6.5cm}{!}{\includegraphics{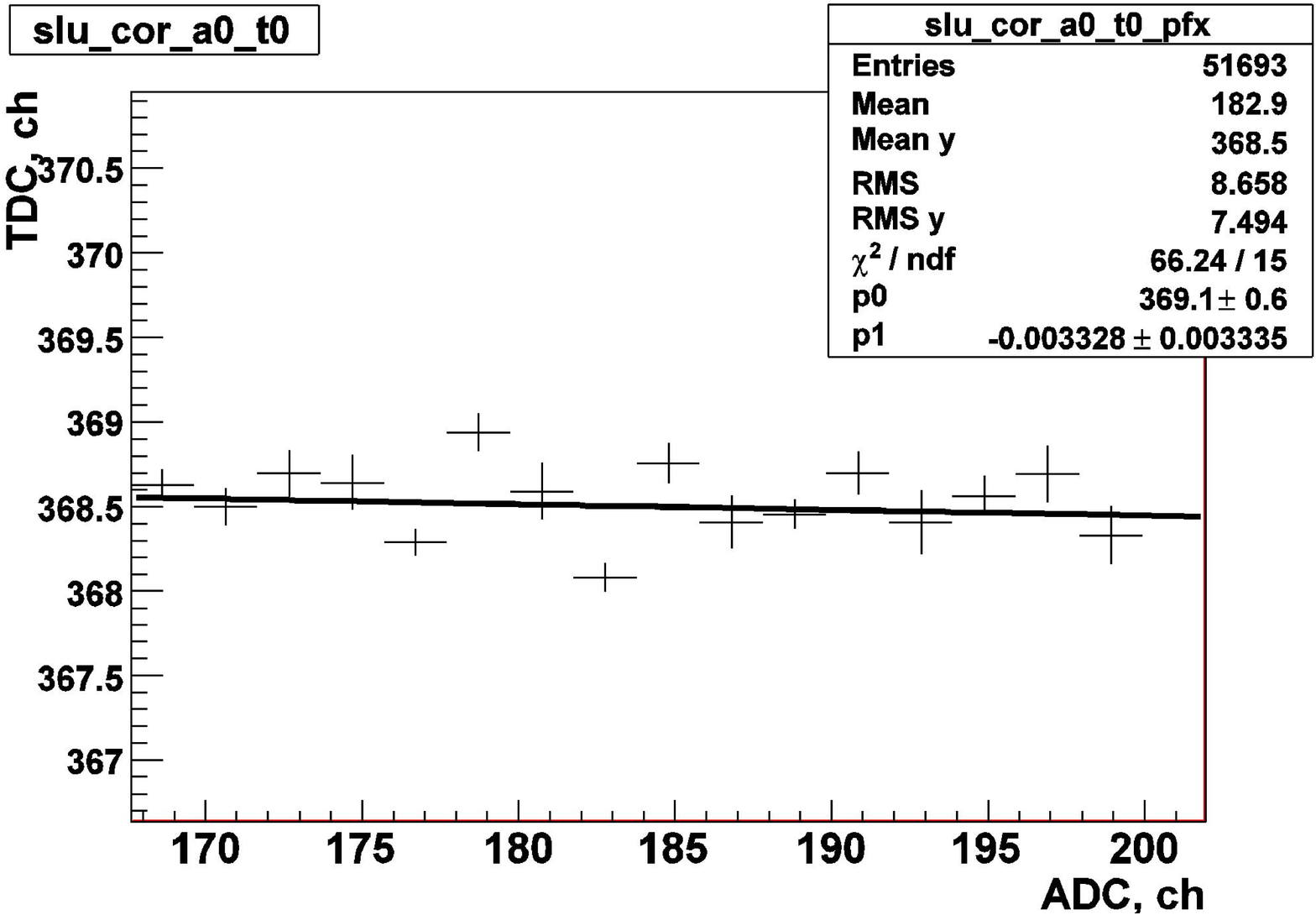}}
 \end{minipage}
\caption{Time-amplitude dependence before (left) and after time-walk  correction (right).
	Solid lines are the data fit by the linear functions.}
\label{fig:fig9}
\end{figure}	 

The timing spectra for all PMT signals  were  corrected taking in account the  time-amplitude    dependence (so called time-walk effect seen in the left panel of Fig.\ref{fig:fig9}). This correction allow to  compensate an amplitude time dependence as shown in the right panel of Fig.\ref{fig:fig9}. 
The final timing resolution for single trigger counter after correction was about 150 ps.

\section{\label{sec:btr} Beam test results.}
 
The light output and timing resolution 
for different combinations of PMT and WLS in the HCAL module prototype have been 
measured with high energy muon beam.
The reference measurements were performed using FEU-84 PMT and WLS of K-7 type. 
It was mentioned above that WLS of K-7 type has maximum in the emission spectrum at 520 nm and decay time about 6 ns. This WLS has a good uniformity and high conversion light output. 
FEU-84 is a 12 stages PMT  with   bi-alkaline   photocathode with diameter 25 mm and maximum of the spectral sensitivity at 420-460 nm.
Such PMT-WLS combination was used previously 
in WA98 \cite{WA98_cal} and COMPASS \cite{Compass_cal} hadron calorimeters.

\begin{figure}[hbtp]
\begin{minipage}[t]{0.47\textwidth}
 \centering
  \resizebox{6.0cm}{!}{\includegraphics{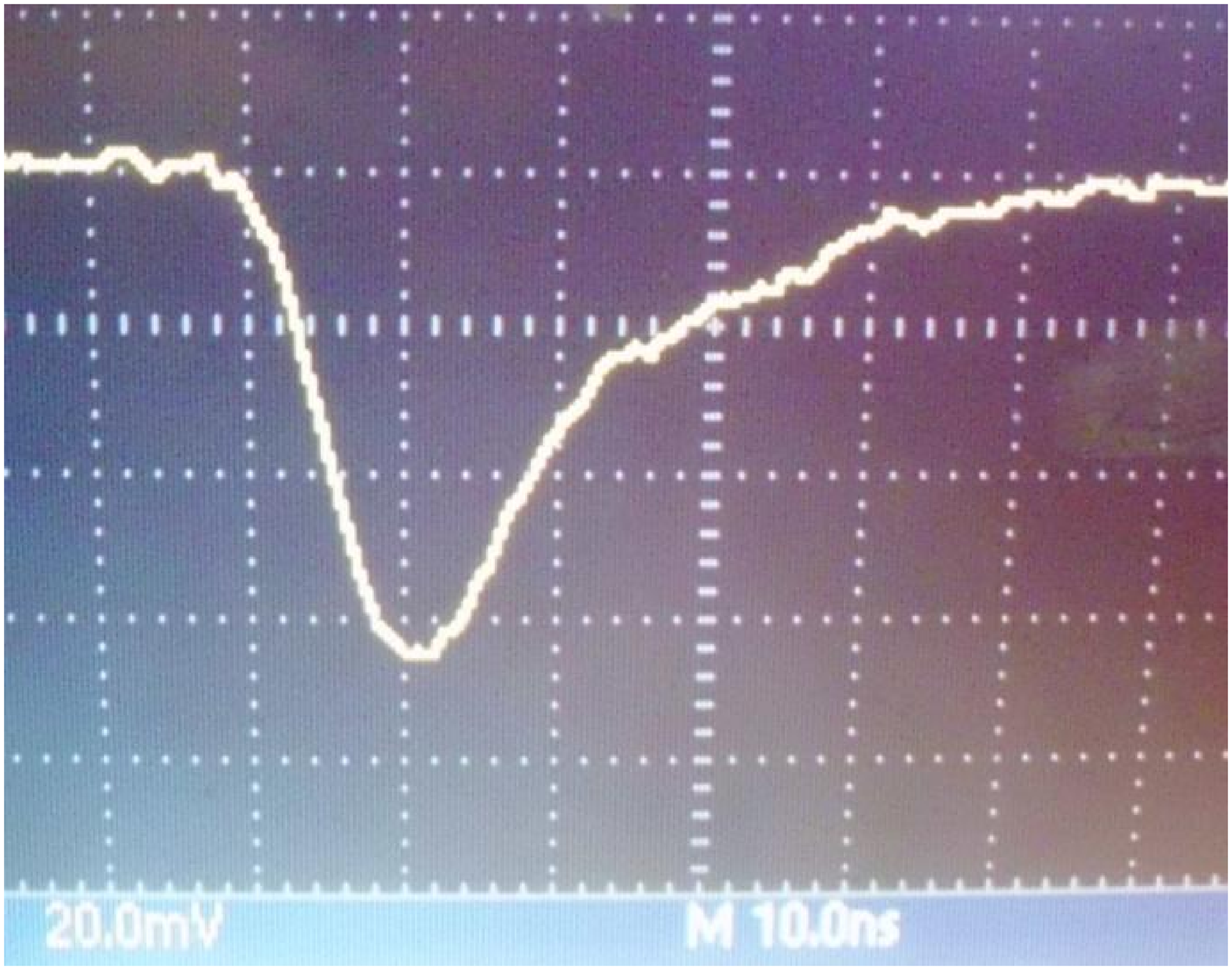}}
\end{minipage}\hfill
\begin{minipage}[t]{0.47\textwidth}
 \centering
  \resizebox{6.5cm}{!}{\includegraphics{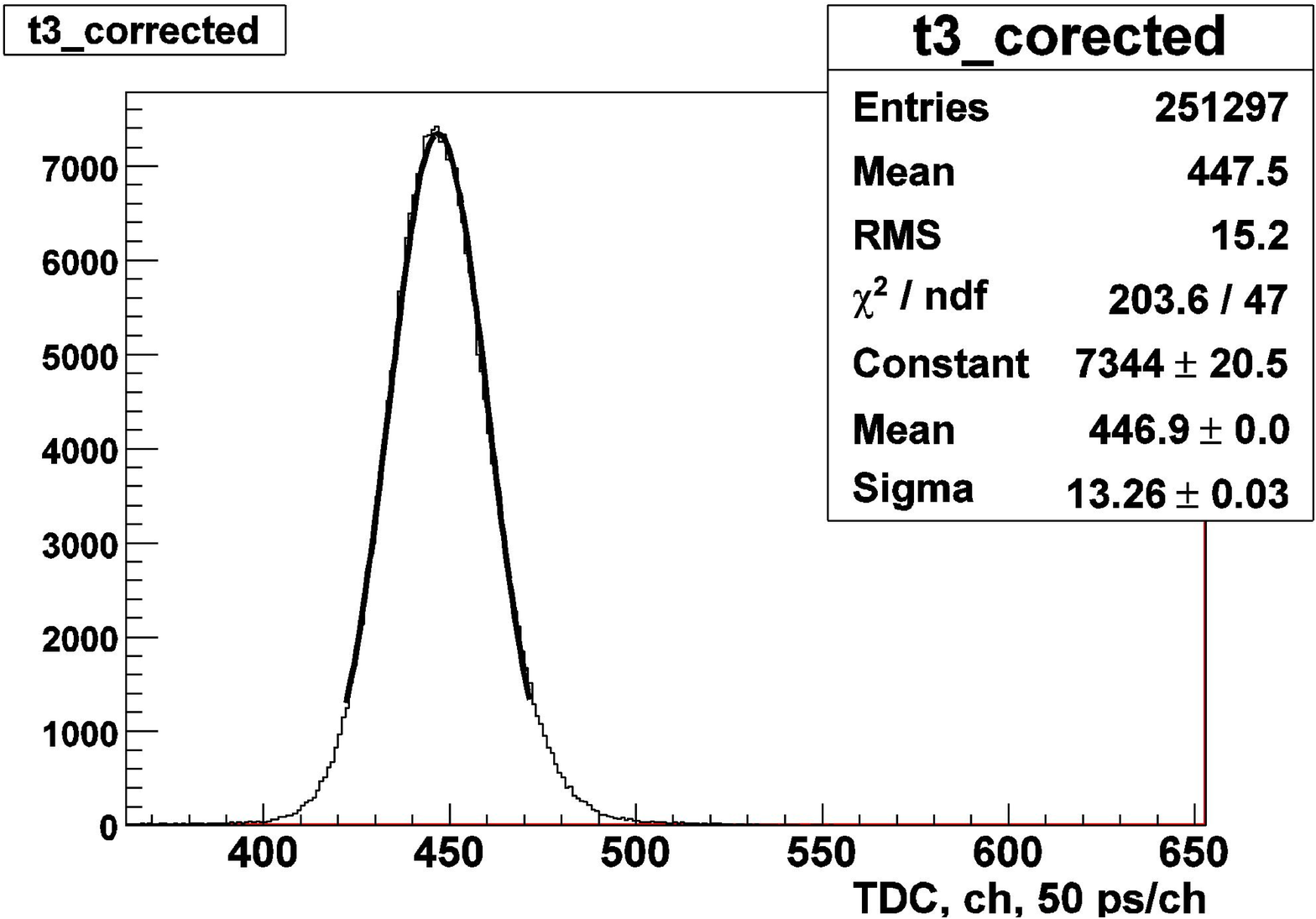}}
 \end{minipage}
\caption{The signal pulse shape for FEU-84 PMT with WLS of K-7 type (left) and its timing spectrum (right) from the high energy muons.
	 Solid line shows the data fit by the Gaussian function. }
\label{fig:fig10}
\end{figure}	 
 
The signal pulse shape for FEU-84 PMT with WLS of K-7 type  
from the  high energy muons passing through the middle of the HCAL module prototype 
is shown in the left panel of Fig.\ref{fig:fig10}. 
The pulse rise time  and its full width at half maximum are $\sim$10 ns and  $\sim$25 ns,
respectively. Timing resolution  of the module prototype  can be obtained from its timing spectrum  presented in the right panel of Fig.\ref{fig:fig10}. It is equal  to $\sim$640 ps taking into account 
the trigger timing resolution of 150 ps. 

EMI-9814KB  is a 14 stage  PMT with 46 mm  bi-alkaline photocathode. The  
pulse rise time  and its full width at half maximum obtained with  high energy muons are $\sim$2 ns and  $\sim$7 ns, respectively,  for the  WLS of the K-7 type used. 
Timing resolution of the HCAL module prototype in this case is  equal to $\sim$400 ps taking into account the trigger timing resolution of 150 ps.

\begin{figure}[hbtp]
\begin{minipage}[t]{0.47\textwidth}
 \centering
  \resizebox{6.5cm}{!}{\includegraphics{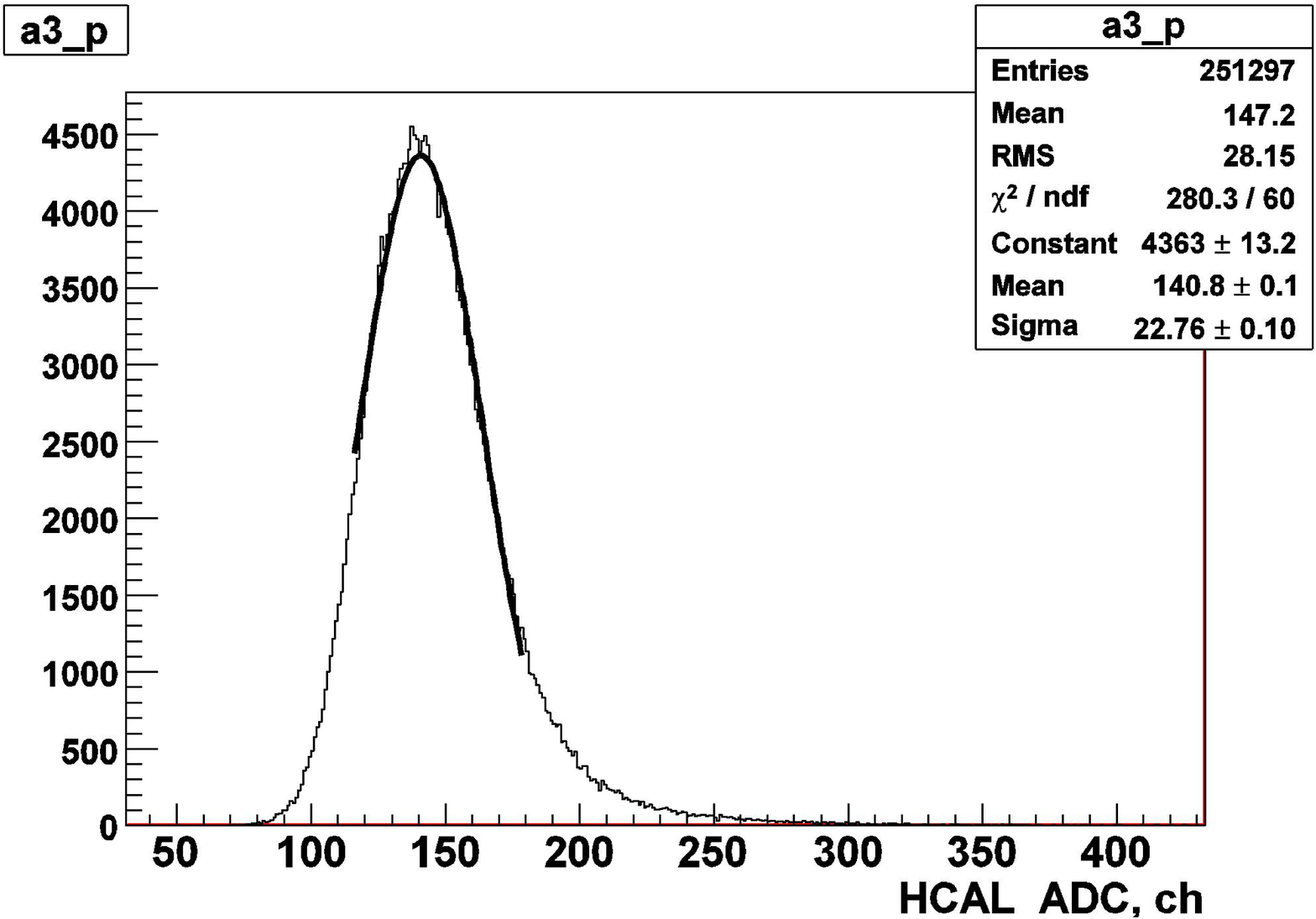}}
\end{minipage}\hfill
\begin{minipage}[t]{0.47\textwidth}
 \centering
  \resizebox{6.5cm}{!}{\includegraphics{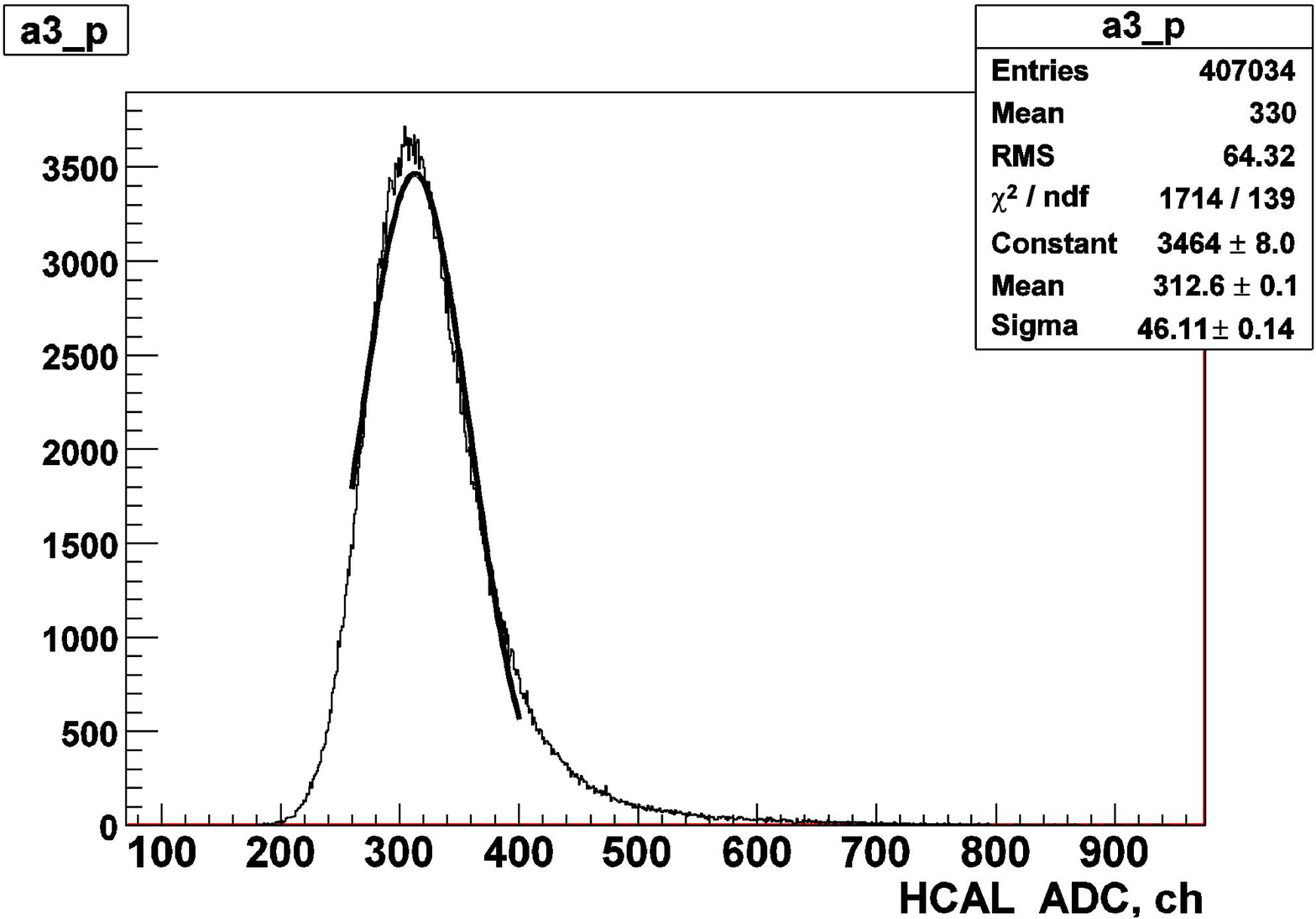}}
 \end{minipage}
\caption{The HCAL module prototype amplitude spectra  from the high energy muons for  FEU-84 (left) and EMI-9814KB (right). The WLS of the K-7 type is used. Solid lines show the data fit by the Gaussian function.}
\label{fig:fig11}
\end{figure}	 

The HCAL module prototype amplitude spectra  from the high energy muons for the  FEU-84 and EMI-9814KB
light detection are presented in the left and right panels of Fig.\ref{fig:fig11},
respectively. The calibration the LeCroy2249A module charge scale was done using LED signal.  
The amplitude signal from the LED has been fitted by the Gaussian function. 
The number of photoelectrons $N_{p.e.}$ has been obtained as $N_{p.e.}=(\bar{A}/\sigma_A)^2$,
where $\bar{A}$ and $\sigma_A$ are the mean value and standard deviation given by fit, respectively.
The values of the calibration coefficients of $k$=0.59$N_{p.e.}$/channel and  $k$=0.49$N_{p.e.}$/channel  were obtained for  FEU-84 and EMI-9814KB, respectively.  
The numbers of photoelectrons corresponding to the amplitude spectra obtained from muon beam and 
presented in  Fig.\ref{fig:fig11} are $N_{p.e.}$=83 and $N_{p.e.}$=152 for FEU-84 and EMI-9814KB, respectively. Therefore, the  light output for  EMI-9814KB is $\sim$1.8 times larger  than for 
FEU-84. 

\begin{table}[hbt!]
\label{table:pmt-wls}
\caption{The light output and timing resolution for HCAL module prototype
for different PMT-WLS combinations.}
\vspace{5mm}
\centering
\begin{tabular}{|c|c|c|c|}
\hline
PMT type & WLS type & Light output, & Timing resolution, \\
         &          & $N_{p.e.}$ & ps\\
\hline
FEU-84 &  K-7 & 83 & 640 \\
EMI-9814KB & K-7& 152 & 400 \\
EMI-9814KB & K-30& 108 & 408 \\
\hline
\end{tabular}
\end{table}

The dependence of the amplitude and timing response of the HCAL module prototype 
on the WLS type has been investigated with EMI-9814KB PMT. 
The similar timing resolution as for WLS of K-7 type has been obtained with WLS of K-30 type, namely, 
408 ps.  However, the light output is $\sim$1.4 times lower than in the case of WLS of K-7 type use.
Further optimization of the  WLS painting procedure is necessary to  obtain better light output with K-30 luminophore. 
The results of these measurements are presented in Table 2.

\begin{table}[hbt!]
\label{table:uniformity}
\caption{The normalized signal amplitude, time appearance and timing resolution for HCAL module prototype
for different $X$ and $Y$ positions of the muon beam impact point. $X$=0, $Y$=0 is the prototype centre.}
\vspace{5mm}
\centering
\begin{tabular}{|c|c|c|c|c|}
\hline
$X$- beam & $Y$- beam  & Light             & Time                  & Timing   \\
position, &  position, & output $\bar{A}$, & appearance $\Delta T$, & resolution $\sigma_T$,  \\
      mm  &    mm      &         \%       & ps & ps \\
\hline 
0 &  40 & 100  & 0  & 410\\
0 & -35 & 94  & 550 & 410\\
35 & 40 & 105  & 50 & 410\\
35 &-35 & 91  & 800 & 410 \\
\hline
\end{tabular}
\end{table}
 
Test of an uniformity of the HCAL module prototype was performed using muon beam 
directed to a given point of the prototype front side. PMT  EMI-9814KB 
and WLS of K-7 type were used for the measurements. The direction of the muon beam was
parallel to the module axis.  The amplitude of the signal, its time appearance and timing resolution of the prototype were measured as a function of the muon beam impact point.   
The results of these measurements are presented in Table 3. 
Here $X$=0 and $Y$=0 correspond to the prototype centre.
The signal amplitudes were normalized to
the signal amplitude obtained for the muon beam hit in the impact point with
the coordinates $X$=0 and $Y$=40. One can see that the normalized amplitude $\bar{A}$ decreases as 
the distance between the muon beam impact point and WLS is increased. 
The difference of the amplitudes for different coordinates of the muon beam 
impact point on the prototype  reflects the light attenuation in the scintillation plates. Maximal amplitude changing is observed in the vertical direction as  $\sim$10-15\%.  
Possible way to  improve the light output uniformity of the module prototype could be 
the scintillator thickness increasing from 4 mm  to 5 mm. The Monte-Carlo simulation
demonstrated that the energy resolution of the calorimeter will not change significantly
in this case.
The signal time appearance  $\Delta T$ increases
as the  distance between the muon beam impact point and WLS is increased, while  
the timing resolution $\sigma_T$ was found to be 410 ps independently  on the muon beam impact point. 
The dependence of the $\Delta T$ on the distance from the hit position to WLS 
will play a role in the case of hadronic shower increasing the  signal duration 
from single HCAL module. The observed uniformity of the  timing resolution $\sigma_T$ across
the section of  HCAL module can indicate that there is no
the ZDC energy resolution decreasing due to hit position dependence.   
However, the systematic studies of the HCAL module with hadron beam is required.

\section{\label{sec:conclusions} Conclusions}

\begin{itemize}
\item
The Monte-Carlo simulation results demonstrated the feasibility to use the proposed ZDC for
the centrality determination in heavy-ion collisions for baryonic matter properties studies
at Nuclotron energies. 
\item 
The results of Monte-Carlo simulation on the energy resolution of the hadron calorimeter in the
energy domain of Nuclotron has been used to prepare HCAL module prototype.
\item
The response of the full scale  prototype of the hadron calorimeter module 
for the peripheral part of BM@N ZDC to high-energy muons has been measured at U-70 accelerator 
for different combinations of PMT and WLS.  The amplitude and timing spectra have been obtained.  
%Measurement of hadron calorimeter module response to high-energy muons shown as amplitudes of signals 
%and time  resolution depended from various combinations of PMT and WLS.
\item
The results demonstrate that the use of EMI-9814KB PMT 
provides better timing resolution and $\sim$1.8 times larger light output for the HCAL module 
prototype than for  FEU- 84 PMT for WLS of K-7 type. 
The use of WLS of K-30 type with EMI-9814KB PMT shows also a good  timing resolution.
However, further  optimization of the surface  coating is required 
to increase light output as a whole.	
\item The non-uniformity  of the light output due to attenuation in the scintillation plates was found $\sim$10-15\%. This value can be improved by the scintillator thickness increasing 
from 4 mm  to 5 mm.
\end{itemize}
\vspace{0.5cm}
The authors are grateful to Dr.~A.Yu.~Isupov for fruitful comments and remarks.
They thank Prof.~K.K.~Gudima for the help in the simulation with LAQGSM generator.


\begin{thebibliography}{99}
\bibitem{bmn_CDR} \textit{Ablyazimov T.~O. et al. (BM@N Collaboration)} // 
Conceptual Design Report of BM@N. 
http://nica.jinr.ru/files/BM@N/BMN-CDR.pdf
\bibitem{nica} \textit{Kekelidze V.~D., Sorin A.~S. et al. (NICA Collaboration)} //
http://nica.jinr.ru
\bibitem{bmn_PoS} \textit{Ladygin V.~P. et al. (BM@N Collaboration)} //
{PoS Baldin-ISHEPP-XXI. 2012. P.038.}
\bibitem{brat1} \textit{Bratkovskaya E. et al.}// Nucl. Phys. A. 2013. V.914. P.387.
\bibitem{vasiliev_npps2011}  \textit{Vasiliev T.~A., Ladygin V.~P. and Malakhov A.~I.} //
{Nucl. Phys. Proc. Suppl. B. } 2011. V.219-220. P.312.
\bibitem{svd2} \textit{Kokoulina E. et al. (SVD-2 Collaboration)} //
PoS ICHEP2012. 2013. P.259.
\bibitem{bmn_dspin2013} \textit{Ladygin V.~P. et al.}//
In Proc. of the XV-th Advanced Research 
Workshop on High Energy Spin Physics (DSPIN-13), 
8-12 October 2013, 
Dubna, Russia; Edited by A.V. Efremov and S.V. Goloskokov, JINR, Dubna,  
ISBN 978-5-9530-0315-3. 2014. P.239.
\bibitem{terekhin_PoS} \textit{Terekhin A.~A. et al.}// 
PoS Baldin-ISHEPP-XXI. 2012. P.005.
\bibitem{piyadin_C12} \textit{Piyadin S.~M. et al.} //
{Phys.Part.Nucl.Lett.} 2012. V.9. P.589.
\bibitem{WA98_cal}
\textit{Arefiev V.~A. et al.}// JINR Rapid Communications. 1996. V.5[79]-96. P.15.
\bibitem{Compass_cal} \textit{Vlasov N.~V.  et al.}// Instrum. Exp. Tech. 2006. V.49. P.41.
\bibitem{balandin} \textit{Balandin V.~P. et al.}// JINR Preprint. R13-2008-196. 2008. Dubna. 
(in Russian).
\bibitem{urqmd}\textit{Bass S.~A. et al.}//
Prog.Part.Nucl.Phys. 1998. V.41. P.225;

\textit{Bleicher M. et al.}// J.Phys. G. 1999. V.25 P.1859.

\bibitem{laqgsm}\textit{Toneev  V.~D.  and Gudima K.~K. }// Nucl.Phys.A. 1983. V.400.  P.173c;

\textit{Toneev V.~D., Amelin N.~S.,  Gudima K.~K. and Sivoklokov S.~Yu. }// Nucl.Phys.A. 1990. V.519. P.463c;

\textit{Amelin N.~S., Staubo E.~F., Csernai L.~S.  et al.}// Phys.Rev.C. 1991. V.44.  P.1541.

\end{thebibliography}
\end{document}